\documentclass[twocolumn,a4paper,aps,pre,preprintnumbers]{revtex4}
\usepackage{graphicx}
\usepackage{amssymb,amsmath,amsthm,subfigure,hyperref}

\newcommand{\comment}[1]{}

\begin{document}
\title{Thermodynamic approach for community discovering within the
  complex networks: LiveJournal study.}
\author{Pavel Zakharov}
\affiliation{Department of Physics, University of Fribourg, CH-1700,
  Switzerland, email: Pavel.Zakharov@unifr.ch}

\date{\today}

\begin{abstract}
The thermodynamic approach of concentration mapping is used to
discover communities in the directional friendship network of LiveJournal
users. We show that this Internet-based social network has a power-law
region in degree distribution with exponent $\gamma = 3.45$. It is
also a small-world network with high clustering of nodes. To study the community structure we
simulate diffusion of a virtual substance immersed in such a network as in
a multi-dimensional porous system. By analyzing concentration profiles at
intermediate stage of the diffusion process the well-interconnected
cliques of users can be identified as nodes with equal values of concentration.
\end{abstract}

\maketitle

\begin{figure}
\centering
  \includegraphics[width=\linewidth]{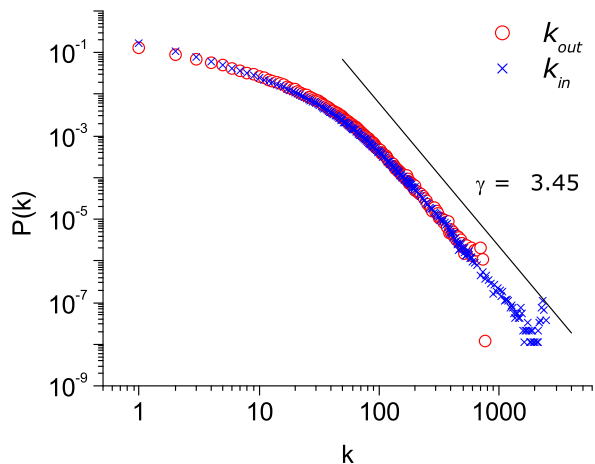}
  \caption{Probability density functions of in- and out-degrees for
  LiveJournal 
  users. The line shows a slope of -3.45 which equally well fits
  $P(k_{in})$ and $P(k_{out})$.}
  \label{fig:degrees}
\end{figure}

\section{INTRODUCTION}

In recent years there has been an enormous breakthrough in research of
complex networks due to the application of statistical
physics methodology \cite{albert:review,dorogovtsev:review,dorogovtsev03:book}. Many different complex
systems instead of being completely random prove to have signatures
of organization such as clustering and power-law distribution of
links. Together with the small-world property \cite{milgram:small} these are the inherent features
of an extremely wide variety of systems such as the World-Wide
Web \cite{albert:diameter,kleinberg:web99,kumar:web00},
Internet \cite{satorras:internet}, collaboration networks of movie
actors \cite{watts:1998,newman:random} and scientists \cite{newman:random}, the web of human
sexual contacts \cite{liljeros:2001} and many others. In spite of the
fact that some concepts of complex networks theory were originally
introduced in sociology the statistical study of social networks is
complicated by the difficulty in reliable data collection due to certain privacy and ethical reasons.
 One of the solutions for this problem is the analysis of collaboration
networks \cite{watts:1998,newman:random}, e-mail
interactions \cite{arenas:email,arenas:community,newman:email}, instant
messaging \cite{smith:2002} and online blogging \cite{kumar:bursty,kumar:structure,nowell:phd,nowell:pnas}.  
\comment{
Nowell {\em et al.} recently studied geographic aspects of LiveJournal
(www.livejournal.com) blog space  and they reported parabolic shape of
friendship degrees distributions \cite{nowell:pnas}. }
Here we studied basic structural properties of LiveJournal blog service
social network and demonstrated the diffusion-motivated method to
discover communities on the case of this network.

\section{LIVEJOURNAL NETWORK}

LiveJournal (LJ) is an online web-based journal service
with an emphasis on users interactions \cite{lj:faq}. In January 2006 it had $9.3 \cdot 10^6$
users in total, $2.0
\cdot 10^6$ of them were {\em active in some way} according to
official  LiveJournal statistics \cite{lj:stat}.
The essential feature of LJ service is the ``friends'' concept which helps
users to organize their reading preferences and 
provides security regulations for their journal entries and personal data. Friends list
is an open information and can be accessed through a conventional WWW
interface or through a dedicated bot interface provided by LJ system.

Data collection was performed by crawler programs running simultaneously
on two computers and exploring the LJ space by following directional
friendship links starting from two users with a large number of
incoming friendship links. For each user the crawler was obtaining his friends list (outgoing links) and the
number of users who have the given user in their friends list (incoming
links). Each user from the friends list which was not yet explored by
the crawler was added to the end of the processing queue if he was not already
there. If the user was in the queue his queue score was
incremented every time he was found in someones' friends list. Users
with higher queue scores were processed first. This ensured fast
collection of the essential part of the network. Basically this algorithm 
is a modification of Tarjan's depth-first search algorithm for
finding the connected component of a graph \cite{tarjan:alg72,hopcroft:alg73}.
Total time of collection was 14 days with the total number of 
discovered users $3\:746\:264$ found in a connected component. We are aware
that during the time of collection the network was 
undergoing continuous changes. We estimated the number of users deleted from the LJ
database but still present in the friends lists was less than 0.1\%
which makes us believe that the evolution of LJ network did not
influence our statistics much.

The estimated probability distribution functions of in- and out-degree are presented in
log-log scale in the Fig.~\ref{fig:degrees}.  The estimated mean
of the numbers of  outgoing  and incoming friendship links is $\langle k_{out}\rangle =
15.91$ and $\langle k_{in}\rangle = 16.07$, correspondingly. The
average in-to-out ratio $\langle k_{in} / k_{out}\rangle =
1.157$. The number of incoming links is slightly larger than the
number of outgoing due to the fact that only the outgoing links were
used for crawler navigation so some of the LJ users were unreachable by
directional links but they were listed in the users pages. 

There are also several technical restrictions for the degrees: maximum
number of friends per user is 750 and only 150 of them can be listed on the users' info
page and can be effortlessly accessed by the LJ users. From our experience LJ bots interface
does have some problems listing the users who consider a certain
user as a friend if there are more than 2500 of them hence we cut the
data at $k_{out\: max} = 2500$. 

 As one can see from the Fig.~\ref{fig:degrees} in- and out-degree distributions
reveal a power-law decay $P(k) \sim k^{-\gamma} $ for $k > 100$ with 
the value of the exponent $\gamma_{in} \approx \gamma_{out} =  3.45 \pm 0.05$ which is
surprisingly close to the values $\gamma_{in} \approx \gamma_{out} \approx
3.4$ obtained by Liljeros {\em et al.} for sexual
contacts \cite{liljeros:2001}. Scaling of the distributions contradicts the
results of Nowell {\em et al.} \cite{nowell:pnas}
who reported parabolic shape of LJ degrees distributions.
The skewness of 
the distributions in our case can be explained by the social origin of
LJ network. As it is pointed out by Jin {\em et
  al.} \cite{jin:structure} degree distribution for social
networks does not appear to follow power-law distribution due to the
cost in terms of time and efforts to support friendship. In the case of
LJ network the cost of friendship is the size of friends feed which
accumulates all the recent entries of the user's friends. We can also
separate two classes of LJ users: ``readers'' and ``writers''. The first are
mainly using their accounts to read the journals of others. They update
journals only episodically and are not deeply involved in LJ
community life. They do not have many incoming and outgoing links and
they are responsible for skewness of the distributions for $k <
100$. Meanwhile active ``writers'', representing minority of the registered users 
exploit full capacity of LJ system. They spend much time participating
in LJ community life, and they have a larger number of incoming and
outgoing links which are distributed by power-law.

The origin of power-law region in the distributions can be explained by
continuous evaluation and self-organization of the LJ network and preferential attachment
mechanism similar to the general WWW growth mechanism
\cite{barabasi:1999}. One an interesting journal gets popular it will
be cited and promoted in 
the journals of its readers which will help to further increase its
popularity which leads to a ``rich-get-richer'' effect occurring
in many network systems
\cite{barabasi:1999,dorogovtsev:review}. However linear growth
with linear preferential attachment protocol leads to a power-law
degree distribution with $\gamma = 3$ which is smaller than the exponent
obtained for our study. Larger values of exponent can be
explained by alternative growth mechanisms: preferential attachment
with rewiring \cite{albert:topology00} and copying mechanism
\cite{kleinberg:web99,kumar:web00}. Rewiring in LJ system implies that
users are not only establishing new friendship links but also breaking
the old ones while copying occurs when the user inherits part of the
friendship connections of his friends. The latter effect is called
''transitivity'' in sociology \cite{wasserman:94} and is responsible for
users cliques formation or clustering.

We characterize clustering of LJ users by calculating the clustering
coefficient as introduced by Watts and Strogatz 
\cite{watts:1998,albert:review}. It is defined as the number of links between user's friends divided
by the maximum possible number of links between them averaged over all
users in the network. If the user $i$ has $k_i$
friends with $E_i$ links between them the maximum possible number of
directed links is $k_i (k_i - 1)$ and the clustering coefficient for the
user $i$ in the case of directed network can be defined as:
\begin{equation}
C_i = \frac{E_i}{k_i (k_i - 1)}.
\label{eq:clustering}
\end{equation}
The average clustering coefficient for the whole network as calculated
from our data is: $C = \langle C_i \rangle_{i=1..N} \approx 0.3302$. It is worth to
compare this value to the clustering coefficient of a random
directional Erd\H{o}s-R\'{e}nyi graph which can be found as $C_{rand}
= \langle k \rangle / (N - 1)$ which for LJ network is ca. $4.24 \cdot 10^{-6}$. The fact that actual 
clustering coefficient for LJ network is nearly five orders of magnitude larger than
it would be expected from randomly linked network with the same degree and
size is a clear indication of high user clustering.

The peculiar feature of the LJ network is the high 
reciprocity \cite{wasserman:94} of friendship links. We found that 79.26\% of links 
are bi-directional which means that this percentage of outgoing links 
is returned as incoming and {\em vice versa}
the same percentage of incoming links originates from users friends.
This value is higher than reciprocity 57\% found for the WWW
\cite{newman:email02} which is the technical environment of LJ. Increasing
of reciprocity may be explained by social origin of LJ network. Due to
the rules of social interactions user $A$ usually feels obliged to establish a
friendship connection to the user $B$ if such a connection was already
established by $B$ to $A$. Another explanation for high reciprocity is
that often relations in  
LJ space is based on real-life people relations which means that
LJ users are linking to the other users which are their friends in the real
world. In this case the LJ network directly inherits the undirectional
structure of the underlying social network.

\begin{figure}
\centering
  \includegraphics[width=\linewidth]{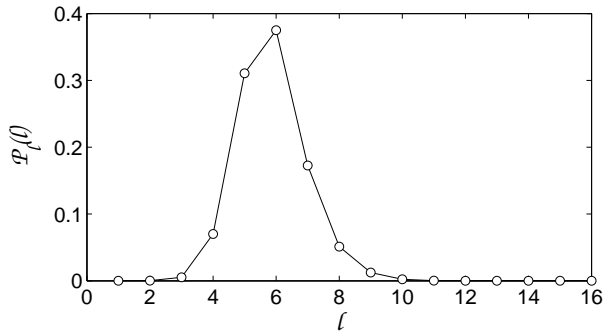}
  \caption{Probability distribution function of the minimum path length
    between LiveJournal users through the directional friends links.}
  \label{fig:path_distance}
\end{figure}

In order to characterize small-world properties of LJ network we estimated the
probability distribution function $P_\ell(\ell)$ of the minimum path distance
or hopcount between the nodes through directional links. The results are presented in
the Fig.~\ref{fig:path_distance}. The average distance estimated for
our set of data is $\langle \ell \rangle = 5.86$. 
\comment{
According to the
general approach developed by Newman \textit{et al.}
\cite{newman:random} an average path length can be estimated using the
following expression:
\begin{equation}
\mathcal{\ell} = \frac{ln(N / z_1)}{ln(z_2/z_1)} + 1,
\end{equation}
where $N$ is the size of the network and $z_1 = \langle k_{out} \rangle $ and $z_2$ is the number
of the first and the second neighbours. From this we obtained $\ell \approx
4.3$ which is significantly smaller than the value obtained from
the distribution. We are considering this as a first sign of structure
within LJ network. 
}
Based on the recently obtained expression for the mean distance between
the nodes in scale-free networks by Hooghiemstra \textit{et al.} \cite{hoog:mean05}
who improved the widely used result of Newman \textit{et
  al.} \cite{newman:random} the value of $\langle \ell \rangle$ can be estimated
as the following:
\begin{equation}
\langle \mathcal{\ell} \rangle_{th} \approx \frac{ln N}{ln \nu} + \frac{1}{2} -
\left ( \frac{\gamma_e + ln \mu - ln (\nu - 1)}{ln \nu} \right ) - 2
\frac{\epsilon}{log \nu},
\label{eq:hoog}
\end{equation}
where $N$ is the size of the network, $\mu = \langle k \rangle$,
$\nu = \langle k (k - 1)\rangle / \langle k \rangle$, $\gamma_e
\approx 0.577$ is the Euler-Mascheroni constant, and $\epsilon$ is the
expectation of the logarithm of the limit of a super-critical
branching process which depends on the scaling exponent $\gamma$ 
and belongs to the half-open interval $(-1,0]$, where the lower
  boundary is the numerical extrapolation of the results from
  \cite{hoog:mean05} and the upper boundary the theoretical prediction
  for $\gamma > 3$.

For LJ data the equation \eqref{eq:hoog} gives the following range of the mean distance:
$ 4.53 \le \langle \mathcal{\ell} \rangle_{th} < 5.05$ which is in any case
smaller than statistically obtained value. This theoretical prediction
assumes the homogeneity of the graph, and  we believe the possible reason
for such an underestimation of the mean path length is the 
macroscopic structuring of the network which is discussed further.

\begin{figure}
\centering
  \includegraphics[width=\linewidth]{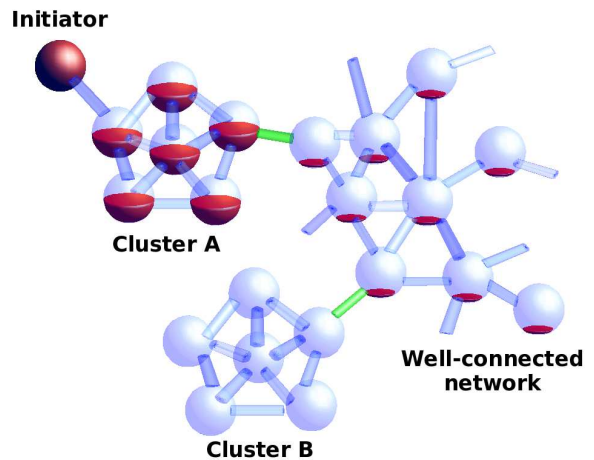}
  \caption{Illustration of the community detection algorithm. After
    diffusion process starts from the initiator node virtual ink
  propagates through network links. Communities can be recognized as the
    groups of nodes with similar amount of ink.}
  \label{fig:scheme}
\end{figure}

\section{COMMUNITY DISCOVERING METHOD}

It seems to be quite natural for the nodes of the complex networks
to aggregate into macroscopic structures with high internal links
density and weak connection to the rest of the network. Such groups are
often referred to as communities. Particular reasons for communities 
formation may depend on the type of the network but this feature
proved to be quite universal and can be found in social, biological and
computer networks \cite{girvan:pnas02,newman:fast}. Finding these
structures within the network is the major step towards understanding its topology.

This problem is known as a graph-partitioning problem in graph theory
and has a nondeterministic polynomial (NP) complexity  which makes it
almost inapplicable for large networks. 

Recent advances in the study of complex networks stimulated the search
of alternative techniques for community discovering and many original solutions
were proposed
\cite{girvan:pnas02,newman:mixing,newman:community,newman:fast,clauset:2004,pons:rw05,wu:physics04,simonsen:diff04}. 
These algorithms can be divided into two main classes: \textit{divisible}, which
hierarchically split the network by removing edges with the highest
betweenness \cite{girvan:pnas02,newman:community} and
\textit{agglomerative} which start from the maximal community
division when each node belongs to its own separate community and
continuously merge these communities basing on some parameter of
nodes similarity \cite{wu:physics04,pons:rw05} or optimizing
the partitioning. In their recent work Clauset \textit{et al.}
\cite{clauset:2004} used the greedy optimization in order to maximize
the \textit{modularity} measure of partitioning quality
\cite{newman:community,newman:fast}. Currently this method is one of the
fastest and runs in time $O (M H ln N)$, where $M = \langle k \rangle N$ is the number
of edges in the network and $H$ is the number of decomposition levels
which is usually small ($H = O (ln N)$)
\cite{clauset:2004,pons:rw05}. In a sparse network the degree is limited
and $M = O (N)$ and so the complexity is $O (N ln^2 N)$ which makes it
fastest nowadays.

Here we propose a method to find communities based oh the principles
of thermodynamics. When the system gets large enough so that the behavior
of its microscopic constituents can be successfully averaged to give
basis for a scientific descriptions of phenomena with avoidance of
microscopic details. Since in thermodynamics behavior of the system can
be described without solving the equation of motion of every
constituent molecule we believe that structure of the large complex
network can be explored without explicit solution of graph
partitioning problem.

Our current study is based on the simulation of a mass diffusion process in the complex network
as in a multi-dimensional porous system with directional links following
physical laws. The diffusion process initiated at one of the nodes by
addition of the virtual ink produces a non-uniform mass distribution at the intermediate state
which can be used to reveal well-interconnected communities within the
complex network by selecting the nodes with similar concentration
values. In this sense our method falls in the class of agglomerative
techniques with the concentration as the similarity measure. However, it
can be shown that the quantity $r_{AB} = | ln \phi_A - ln \phi_B |$,
where $\phi_A$ and $\phi_B$ are two values of concentration  in the nodes $A$
and $B$, as the measure of distance between these nodes. Thus
edge betweenness, characterized as the drop of the logarithm of concentration
along the edge, can be used for hierarchical decomposition of the
network.

The similar measure of distance between nodes based on the random walk
has been recently introduced by Pons \textit{et al.} \cite{pons:rw05}
for the class of undirected networks. It is defined as the difference in probabilities
for a random walker to reach nodes the $A$ 
and $B$ in certain number of steps $t$ starting from some node
$Z$. As these probabilities for a large $t$ are mainly determined by
the in-degrees of the nodes the values of distance should be normalized
A short number of steps $t$ may depend on a particular
network and should be known in advance. Pons \textit{et al.} also pointed out
conceptual difficulties of the random walk scheme application for the directed
networks \cite{pons:rw05}. Several other diffusion motivated
approaches proposed recently (\textit{e.g.}
\cite{wu:physics04,simonsen:diff04,fouss:novel05}) are more or
less consistent with random-walk analogy.

In our model we break the similarity with classical random
walks and the theory of flows in the graph \cite{diestel:graph} in
favour of a realistic physical picture. First, we allow nodes to
accumulate substance by assigning to them infinite maximum capacity. 
The direct flow from the node $A$ to the node $B$ is possible if there is
a directed link from $A$ to $B$ and $\phi_A > \phi_B$. The flow rate
in this case depends on the concentration difference $\phi_B - \phi_A > 0$ and the out degree
$k_{out}$ of the node $A$. In the case of $ A < B$ no mass is
delivered directly from $A$ to $B$. Such rules in the limit of
infinite time lead to equilibrium state with equal mass distribution
which meets the physical expectations. 

Network links in our realization represent pipes (Fig.~\ref{fig:scheme}), directed
links  act as pipes allowing mass to pass in one direction. Mass
propagation within the network system is driven by Flick's law of diffusion:
\begin{equation}
dM = - D \frac{\delta \phi}{\delta x} dS dt, 
\end{equation}
where $dM$ is mass change, $\delta \phi / \delta x$ is
concentration gradient and $dS$ is an area element.

For our discrete system this implies that the rate of mass
exchange between the neighbouring nodes is proportional to the difference of masses in these
nodes. Every node uses its outgoing links to deliver mass to its neighbors with
a smaller amount of ink. The amount of ink $\Delta_{out} M_i$
delivered by the node to its $i$th neighbour is:
\begin{equation}
\Delta_{out} M_i = - \frac{\alpha}{k_{out}} (M_0 - M_i),
\label{eq:main}
\end{equation}
where $M_0 > M_i$ and $\alpha$ is the coefficient determining the
transfer rate and is constant for all
nodes. We analyze the mass $M$ contained in the node instead of
the concentration $\phi$ assuming that all nodes have the same
geometrical volume.
The total delivered mass for a node is the following:
\begin{multline}
\Delta_{out} M = \sum_{i=1}^{k_{out}} \Delta_{out} M_i =  - \alpha \left ( M_0 - \frac{1}{k_{out}}
\sum_{i=1}^{k_{out}} M_i \right) =  \\
- \alpha ( M_0 - \overline{M} ),
\end{multline}
where $\overline{M}$ is the mean ink mass in the neighbouring
nodes with smaller masses. Mass transfer in the pipe happens
instantaneously.
 Thus we can apply mass conservation
law and increase mass in the neighbouring nodes by the amount
taken from the node:
\begin{eqnarray}
\Delta_{out} M & = & - \sum_{i=1}^{k_{out}} \Delta_{in} M_i \\
\Delta_{in} M  & = & - \sum_{i=1}^{k_{in}} \Delta_{out} M_i 
\end{eqnarray}

The total change of mass at a certain node is composed of the loss of mass due
to diffusion to the neighbours through outgoing links and gain of mass
by the
amount delivered from neighbors through
incoming links: $\Delta M = \Delta_{in} M + \Delta_{out} M$. This
conservation law is the extension of Kirchhoff's
law \cite{diestel:graph} for the node with non-zero capacity.

In order to prevent inequality due to sequential nodes processing, mass changes
for all nodes were calculated without actual changing the masses and then
values of the masses in all nodes were updated. For the special case of absence of outgoing
links $\Delta_{out} M = 0$ the specific node acts as a virtual ink absorber which can
only gain ink from the neighbours but does not have ways to deliver it
back. Nodes without incoming links are not 
considered due to their invisibility for the data collecting crawler and
thus are absent in our database.

We start by putting an initial amount of ink of $M_0 = N$ mass
 units in one of the nodes which we call the \textit{initiator}. Subsequently system is
allowed to proceed to the equilibrium state by continuous mass
redistribution within the network according to our rules. The
 expectation for an 
equilibrium state for a connected network system is equal
distribution of mass $M_0$ among the nodes so that each of 
them ends up having $M_0 / N = 1$ mass units. While evolving to this state the system
passes through non-equilibrium states with non-uniform mass
distributions.

Imagine a cluster of well connected nodes inside the network
connected to the outside world only by few outgoing and
incoming links. The ink diffusion inside the cluster is relatively fast due to the
presence of a large number of exchange channels between the 
members and a high conductivity of the channels
ensemble. Limited number of channels going outside the cluster forms the
bottleneck for mass delivery. Under these conditions the flow rate between
the members is much higher than between the members and non-members and dispersed ink will
likely form an equi-concentrational \comment{Phys. Rev. E 49, 5431--5437
  (1994)} volume within the cluster. 
Each cluster in this system with
specific connection properties such as flow rate and distance from
the initiator would have in each of its
nodes the same concentration of ink with the value specific to the 
particular cluster. Thus by estimating the probability distribution
function of concentration one can analyze non-uniformity of ink
distribution and reveal separated clusters by determining the
signatures of equi-concentration volumes.

\begin{figure}
\centering
  \includegraphics[width=1.05\linewidth]{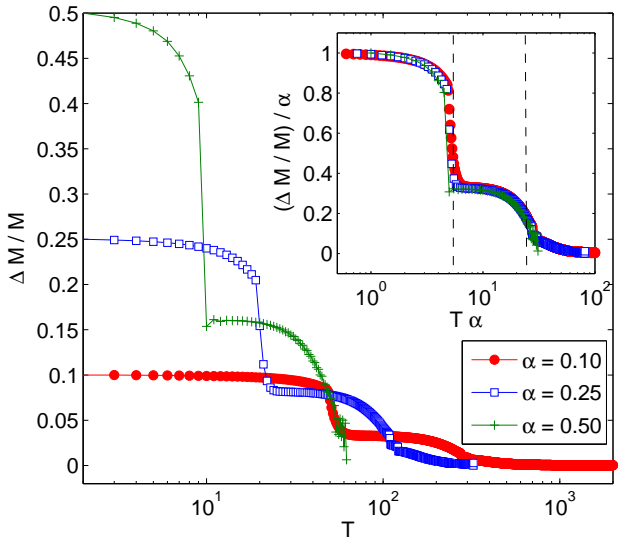}
  \caption{Dynamics of relative  concentration change in the
  initiator node {\em doctor\_livsy} for different flow rates
  $\alpha$. Inset shows rescaled data. Oscillatory parts were cut away.}
  \label{fig:concentration_decay}
\end{figure}

The flow rate $\alpha$ from the equation \ref{eq:main} can be selected
from the half-interval (0;1] and defines the speed of
  simulation. Values larger than 0.5 are not desirable because they
  can cause concentration waves or back-reflections in some cases.

The proposed method does not aim to decompose the whole network on
minimal clusters but to reveal significant clusters within the
network. As we regard the network as an open system which does
not have to be fully described by existing database we do not assign
measure of clustering of the whole network like modularity proposed by Newman
\cite{newman:mixing,newman:community}. However we can quantify the
isolation of the individual community $i$ by parameter of 
confinement $K_i$ which is the characterization of assortative mixing of
individual community. We can define $K_i$ using notation of
Newman \cite{newman:mixing} as following:
\begin{equation}
K_i = \frac{e_{ii}}{\sum_j e_{ij}} = \frac{e_{ii}}{b_i},
\end{equation}
where $e_{ij}$ is the fraction of network edges connecting nodes of
the community $i$ to the community $j$ and $\sum_j e_{ij} = b_i$ is
the fraction of edges starting from the members of $i$. Thus 
parameter $K_i$ defines the number of links connecting the nodes
inside the community $i$ as a fraction  of the total number of links
originating from the members of $i$.

\begin{figure}
\centering
\includegraphics[width=\linewidth]{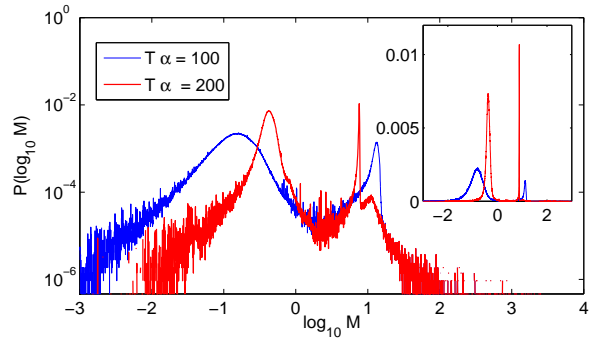}
\caption{Probability distribution functions of virtual ink
  concentration $M$ at two stages of the diffusion process with $\alpha =
  0.1$ and {\em doctor\_livsy} as the initiator node. Inset represents
  the same data in linear scale. Two well pronounced peaks of two
  separated communities are clearly seen.}
 \label{fig:profiles}
\end{figure}

\begin{figure}
\centering
\includegraphics[width=\linewidth]{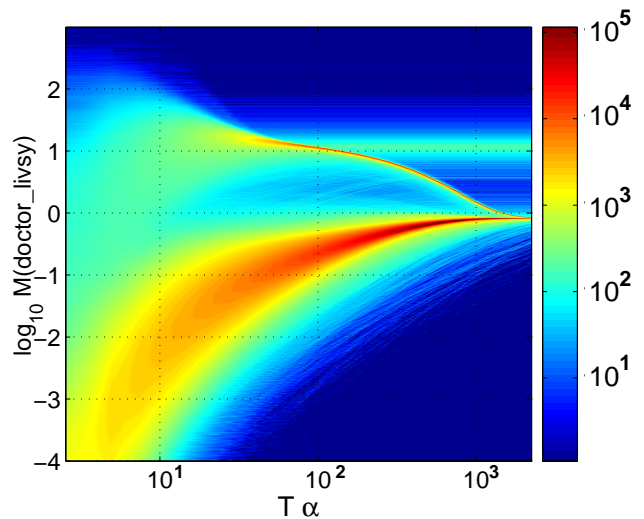}
    \caption{Dynamics of virtual ink distribution within LJ
      network as a logarithmically color coded probability
      distribution function of the ink concentration (vertical 
      axis) and simulation step (horizontal axis). Separation of
      Russian-speaking community (thin upper line, high concentration
      values) from general English-speaking (thicker lower line, lower
      concentration values) can be clearly seen.}
    \label{fig:dynamics}
\end{figure}

\section{RESULTS AND DISCUSSION}

To test our method we performed ink diffusion simulations using our
LJ database starting from different initiator nodes.
Fig.~\ref{fig:concentration_decay} shows the relative mass decay as a 
function of simulation step number $T$ for the flow rates $\alpha =
0.1$, 0.25 and 0.5. User {\em doctor\_livsy} with a high number
of incoming links was chosen as the initiator node. As we will show later
this user belongs to extremely confined Russian-speaking community.
The inset of Fig.~\ref{fig:concentration_decay} shows 
the same data rescaled with respect 
to $\alpha$. As one can see from the match of rescaled curves the
dynamics of the process does not depend on the flow rate $\alpha$ in this
range. The striking feature of the presented data is the obvious
step-like form of the curves which is the effect of non-homogeneous
structure of the LJ network. Flat parts of the $\Delta M / M$ curves
correspond to the exponential decays of $M$ which is the
sign of non-restricted diffusion of ink. The first significant drop of the
decay rate happens when $T \alpha \approx 5$ which is equal to the
double radius of the community to which our initiator belongs. This
corresponds to the moment when virtual ink 
fills the whole community and further expansion of filled area is
impeded by the limited number of links going outside the community.
So if it takes $T_0$ simulation steps for the virtual ink to reach the
borders of the community it also takes $T_0$ simulation steps for the
decay of concentration gradient to reach the initiator node and together this
gives double size of the community.
The second drop at $T \alpha \approx 22$ is not well pronounced and
 corresponds to the filling of the whole network. 

As our community discovering algorithm is based on the detection of
equi-concentration volumes we performed the calculation of the
probability distribution function of $M$ at two stages of
virtual ink diffusion for $\alpha = 0.1$ (Fig.\ref{fig:profiles}). One
can see two well 
pronounced peaks on all plots which occurred to be the Russian speaking
community (larger values of mass $M$) and the rest of LJ network (broader peak at
smaller values of $M$). 

The dynamics of virtual ink distribution is presented in
the Fig.\ref{fig:dynamics}. As it can be seen a distinct separation of the
Russian community peak from the main peak is formed before step
$T \alpha = 50$. At the latter stage it is quite stable and easily distinguishable up to
iteration $T \alpha = 10^3$ which gives quite a long quasi-stationary stage
that can be used for communities detection. It also demonstrates that the
process of equi-concentrational volumes formation is much faster than the
relaxation of the whole system. 

If the initiator node is selected somewhere outside the community the
splitting of the distribution peak is also observed but for this case
average concentration within the Russian community is smaller compared
to the
rest of the LJ nodes. This supports the expectations that if the
community has a limited number of outgoing links it also lacks 
incoming links.

\begin{figure}
\centering
\includegraphics[width=\linewidth]{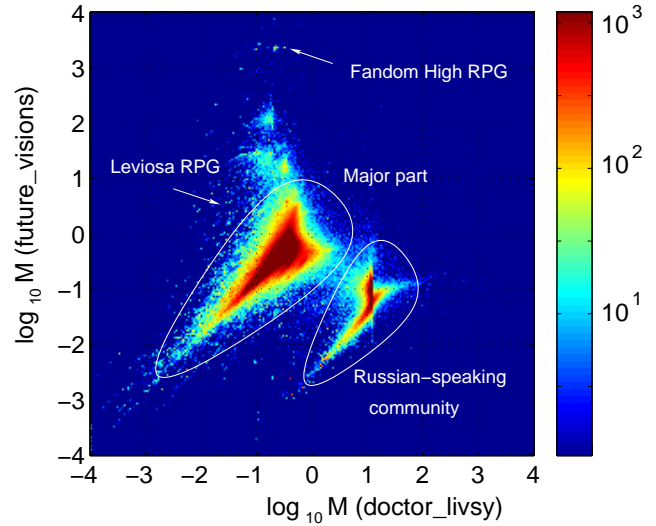}
    \caption{Two-dimensional map of LJ users network obtained by
    concentration configurations of independent diffusion processes
    from two initiator nodes on the stage $T \alpha = 100$.}
    \label{fig:mapping}
\end{figure}

\begin{table*}[t]
\caption{Examples of discovered communities within LiveJournal
  userspace.}
\label{tab:comms}
\begin{ruledtabular}
\begin{tabular}{ccccl}
Representing node & Number of users & Specificity & Confinement $K$ &
Comments \\
\hline
{\em doctor\_livsy} & 227314 & 99.89\% & 98.34\% & Russian speaking
community\footnote{92\% of users have Cyrillic letters in their
  information pages or journals} \\
{\em future\_visions} & 421 & 98.36\% & 96.22\% & Fandom High Role-Playing
Game community \\
{\em alected } & 262 &  99.21\% & 99.10\%  & Leviosa Role-Playing Game community \\
\end{tabular}
\end{ruledtabular}
\end{table*}

The accuracy of community discovering scheme can be improved by
simultaneous simulation of the diffusion from two or more initiator 
nodes. Here we assigned two  independent concentration values to a
single node. All diffusion processes proceed without
influencing each other. The LJ network can now be mapped as a 
probability distribution function of two concentrations and thus the
community can be localized on a two dimensional plot 
as shown in the Fig.~\ref{fig:mapping} for {\em doctor\_livsy} and
{\em future\_visions} as the initiator nodes. One can see two main separated
peaks corresponding to the major part of LJ network and the Russian-speaking
community. The abundance of noise-like spots on the map corresponds to
the small well-separated and well linked communities existing in the
network which are well localized.  

The selection of nodes from a certain community can be performed by simple
thresholding the values of both concentrations. The group of nodes with the
concentration values within the selected range which form the  connected
component in the network can be identified as the community.
The ratio of the number of connected nodes to the total number of
users with concentrations within the range defines the
\textit{specificity} of the method. 

As the complete analysis of LJ community structure as well as the
reasons of their formation is out of the scope of the current paper we
will not list all user cliques found. However in the Tab.~\ref{tab:comms} we
list the largest LJ community and two smaller
ones together with their parameters. The size of discovered
Russian-speaking community is of the order of the total number of LJ 
users from the Russian Federation according to LJ database statistics \cite{lj:stat}
($232\;241$ users in January 2006). The obvious reason for the separation of
this community with a very high value of confinement $K = 98.34$\% is
the prevailing usage
of Russian language. We found by separate analysis of info pages and
journal entries that 92\% of the users within this community are using
Cyrillic alphabet. The fact that the Russian LJ community differs from
the rest of LJ network has been already
pointed out by Internet observers (e.g. Ref.~\cite{gorny:RLJ}).
The two other listed communities are the examples of surprisingly popular class of Role-Playing Game
 communities formed by the virtual users playing characters and 
writing their journals on behalf of these characters.  

\section{CONCLUSIONS}

The LiveJournal friendship network was studied with the general approach
developed for the complex networks and a power-law tail with exponent
$\gamma = 3.45$ was found in the degree distributions. This network
also demonstrated small-world property and high clustering.

To study the community structure we utilized the original thermodynamic approach.
We found that diffusion in an essentially non-euclidean geometry
of a complex network with community structure leads to a peculiar
phenomenon of formation of quasi-stationary equi-concentration volumes
as shown by our simulation. This proves to be very useful
for the detection of well-interconnected groups of nodes. With a limited number of
parallel diffusion processes sufficient for a rough decomposition our method has an $O(N ln
N)$ complexity  (each simulation step analyzes
$M = \langle k \rangle N$ edges which for a sparse matrix $M = O(N)$
and the required number of steps is proportional to the 
diameter of the network which is $O (ln N) $). It is currently one of
the fastest algorithms and was applied for a huge directed network of LJ users 
containing several millions of nodes. To obtain results presented in this 
paper it takes only one or two hours of 
desktop computer time. Moreover this method can be applied locally to
a specific part of the network even with the lack of complete information about distant
parts of the network. The sensitivity of decomposition can be tuned by
increasing the number of initiator nodes with the limit of complete decomposition
when every node acts like initiator of its own diffusion process.

\acknowledgments

Financial support by the Swiss National Science Foundation is
gratefully acknowledged. We thank Frank Scheffold for helpful
discussion. 

\bibliography{/usr/share/texmf/bibtex/bib/base/full}
\end{document}